\documentclass[12pt,a4paper]{article}
\pdfoutput=1
\usepackage{jcappub}
\usepackage{euscript,epsfig,amsmath,amssymb,bm}
\usepackage{amsfonts,latexsym}
\usepackage[normalem]{ulem}
\usepackage{color}
\usepackage{graphicx}

\newcommand{\etal}{{\em et al.}}

\begin{document}

\title{High energy photon polarimeter for astrophysics}

\author[a]{Maxim Eingorn,}
\author[a]{Lakma Fernando,}
\author[a,1]{Branislav Vlahovic\note{Corresponding author},}

\author[b,c]{Cosmin Ilie,}

\author[d,2]{Bogdan Wojtsekhowski\note{Corresponding author},}

\author[e]{Guido Maria Urciuoli,}
\author[e]{Fulvio De Persio,}
\author[e,f]{Franco Meddi,}

\author[g]{Vladimir Nelyubin}

\affiliation[a]{Physics Department, North Carolina Central University, 1801 Fayetteville St., Durham, NC 27707, USA}

\affiliation[b]{Department of Physics and Astronomy, Colgate University, 13 Oak Dr., Hamilton, NY 13346, USA}

\affiliation[c]{Department of Theoretical Physics, National Institute for Physics and Nuclear Engineering, Magurele, P.O.Box M.G.-6, Romania}

\affiliation[d]{Thomas Jefferson National Accelerator Facility, Newport News, VA 23606, USA}

\affiliation[e]{INFN, Sezione di Roma,  I-00185 Rome, Italy}

\affiliation[f]{Department of Physics, Sapienza Universit\`a di Roma, I-00185 Rome,  Italy}

\affiliation[g]{Physics Department, University of Virginia, Charlottesville, VA 22904, USA}

\emailAdd{vlahovic@nccu.edu} \emailAdd{bogdanw@jlab.org}

\abstract{A high-energy photon polarimeter for astrophysics studies in the energy range from 10~MeV to 800~MeV is considered. The proposed concept uses a stack of
silicon micro-strip detectors where they play the roles of both a converter and a tracker. The purpose of this paper is to outline the parameters of such a
polarimeter and to estimate the productivity of measurements. Our study supported by a Monte Carlo simulation shows that with a one-year observation period the
polarimeter will provide 6\% accuracy of the polarization degree for photon energies above 100~MeV, which would be a significant advance relative to the currently
explored energy range of a few MeV. The proposed polarimeter design could easily be adjusted to the  specific photon energy range to maximize efficiency if
needed.}


\maketitle

\flushbottom

\section{Introduction}

Recent discoveries have underlined a key role of astrophysics in the study of nature. In this paper we are presenting a potential instrument for measuring high
energy photon polarization with a proven detector technique which should allow preparation of a reliable tool for a future space-borne observatory.

Polarization of the photon has played an important role (sometimes even before it was recognized) in physics discoveries such as the famous Young's interference
experiment~\cite{Young1802}, Michelson-Morley's test of the Ether theory~\cite{Michelson1887}, determination of the neutral pion parity~\cite{Yang1950,Berlin1950} and many
others, including more recently the spin structure of the nucleon~\cite{Aidala2013}. Polarization of the Cosmic Microwave Background (CMB) will likely be a
crucial observable for the inflation theory (see PLANCK [sci.esa.int/planck] results for constraints on inflationary models based on non-detection of B-modes
and~\cite{Kamionkowski2015} for a review).

During the last decade, observations from the AGILE [agile.rm.iasf.cnr.it] and FERMI-LAT [www-glast.stanford.edu] pair production telescopes have enhanced our
understanding of gamma ($\gamma$) ray astronomy. With the help of these telescopes numerous high energy $\gamma$ ray sources have been observed. However, the
current measurements are insufficient to fully understand the physics mechanism of such $\gamma$ ray sources as gamma-ray bursts (GRBs), active galactic nuclei
(AGNs), blazars, pulsars, and supernova remnants (SNRs). Even though both telescopes cover a wide range of energy (from 20~MeV to more than 300~GeV), neither of
them is capable of sufficiently accurate polarization measurements (recently the authors of~\cite{Giomi2016} estimated that a $5\sigma$ minimum detectable
polarization of $\sim30-50\%$ could be detected by FERMI-LAT for the brightest $\gamma$ ray sources only after 10 years of observation) that could shed light on
the numerous open problems in medium to high energy $\gamma$ ray astrophysics. For example, precise measurements of ray polarization for $\gamma$ rays could be
used, in principle, to determine the nature of the emission mechanisms responsible for blazars, GRBs, X-ray binaries, pulsars, and magnetars. Moreover, measurements
of $\gamma$ ray polarization from distant sources such as blazars or GRBs can help address problems in fundamental physics, such as Lorenz
invariance~\cite{Laurent2011}.

There are several medium to high energy photon polarimeters for astrophysics proposed in the literature. For example, both the NASA group~\cite{Bloser2006} and
the Ecole Polytechnique group~\cite{Bernard2012, Bernard2012a} are considering Ar(Xe)-based gas-filled detectors: the Time Projection Chamber with a micro-well or
micromega section for amplification of ionization. Similarly, the Advanced Energetic Pair telescope (AdEPT), proposed in~\cite{Hunter2013}, is a pair production
telescope capable of polarization measurements using a Three-Dimensional Track Imager based on a low density gaseous (Ar) time projection chamber. Most recently,
e-Astrogam~\cite{DeAngelis2016} has been proposed as a $\gamma$ ray space instrument for the fifth Medium-size mission of  the European Space Agency. In contrast
to the proposals previously mentioned, by the NASA and Ecole Polytechnique groups, e-Astrogam is based on double sided Silicone strip detectors (DSSDs), allowing
high resolution and good sensitivity to linear polarization over its entire bandwidth, from $~0.15$~MeV to $3$~GeV. In this paper we evaluate the features of an
electron-positron pair polarimeter for the full energy range from 20~MeV to 1000~MeV and then propose a specific design for a polarimeter in the 100 to 300~MeV
energy range using silicon micro-strip detectors, MSDs, whose principal advantage with respect to the gas-based TPC is that the spatial and two-track resolution
is about five to ten times better.

The paper is organized in the following way: In section~\ref{sec_moti} we briefly discuss the motivation for cosmic $\gamma$ ray polarimetry in the high energy
region. Section \ref{sec_pol} is devoted to measurement techniques, polarimeters being built and current proposals. In section \ref{sec_flux} we calculate the
photon flux coming from the Crab Pulsar and Crab Nebula. The design of the new polarimeter and its performance are discussed in the last few sections.

\section{Scientific Motivation}\label{sec_moti}
There are several recent reviews of photon polarimetry in astrophysics~\cite{Lei1997, McConnell2004, McConnell2006, Krawczynski2011, Hajdas2013, Pohl2014,
Knodlseder2016} which address many questions open for decades that could find a solution via $\gamma$ ray polarization measurements. We will briefly discuss those
problems in this section. Photon polarimetry for energy below a few~MeV is a very active field of astrophysical research, and some examples of the productive use
of polarimetry at these energies include: detection of exoplanets, analysis of chemical composition of planetary atmosphere, and investigation of interstellar
matter, quasar jets and solar flares. However, no polarization measurements are available in the medium and high energy regions ($\sim1$~MeV -- $\sim1$~GeV)
because of the instrumental challenges.

The primary motivation in proposing a polarimeter is our interest in understanding the emission and production mechanisms for polarized $\gamma$ rays in pulsars,
GRBs, and AGNs by measuring polarization of cosmic $\gamma$  rays in  this under-explored energy region ($\sim100$ -- $1000$~MeV). Additionally, the polarization
observations from the rotation-powered and accretion-powered pulsar radiation mechanisms could help to confirm the identification of black hole
candidates~\cite{Dovciak2008}.

Polarization measurements could reveal one of the possible effects induced by quantum gravity, the presence of small, but potentially detectable, Lorentz or CPT
violating terms in the effective field theory. These terms lead to a macroscopic birefringence effect of the vacuum (see~\cite{Jacobson2006} for more
information). Up to now, the highest energy linear polarization measurement has been for GRB 061122 in the 250-800~keV energy range~\cite{Gotz2014}, and vacuum
birefringence has not been observed in that region. Therefore, extending polarization sensitivity to higher energies could lead to detection of vacuum
birefringence, which would have an extraordinary impact on fundamental physics, or in the case of null detection we could significantly improve the present limits
on the Lorentz Invariance Violation parameter.

Further, according to the observations by the Energetic Gamma Ray Experiment Telescope (EGRET) [heasarc.gsfc.nasa.gov/docs/cgro/egret], the synchrotron emission
of the Crab Nebula is significant in the energy below $\sim$~200~MeV~\cite{Jager1996}.
Additionally, the theoretical studies state that most of the $\gamma$ rays
coming from the Crab Nebula around 100~MeV may come from its inner knot~\cite{Komissarov2011}, so the observations in the neighborhood of 100~MeV will help to
test this theoretical hypothesis and confirm the emission mechanism. In 2007 AGILE and Fermi detected strong $\gamma$ ray flares originating from the Crab system,
with a precise localization and physical origin yet to be determined.
Future polarization measurements will be crucial in answering those two questions, as they
can be used to determine the morphology of the region where particle acceleration takes place.

It is also worth mentioning that polarimetry could test the theories assuming existence of axions (hypothetical particles introduced to solve the strong CP
problem of QCD).
It is interesting that the same axions or axion-like particles can serve as a foundation for a relevant mechanism of Sun luminosity \cite{Rusov2014}. A
theoretical study~\cite{Rubbia2008} has shown that polarization observations from GRBs can be used to constrain the axion-photon coupling: $g_{a\gamma\gamma}\leq
2.2\times10^{-11}~GeV^{-1}$ for the axion mass $10^{-3}$~eV.
The limit on the coupling scales is $\propto 1/\sqrt{E}$; therefore, the polarimetry of GRBs at
higher energies would lead to tighter constraints.

In two of the following subsections we will briefly explain how polarization measurements are involved in confirming the emission mechanism and geometry of two
above-mentioned sources.

\subsection{Pulsars}\label{sec_pulsar}
Pulsars are a type of neutron star, yet they are highly magnetized and rotate at enormous speeds. The questions concerning the way magnetic and electric fields
are oriented, how particles are accelerated and how the energy is converted into radio and $\gamma$ rays in pulsars are still not fully answered. Because of the
extreme conditions in pulsars' interiors, they can be used to understand  poorly known properties of superdense, strongly magnetized, and superconducting
matter~\cite{Guo2014, Buballa2014}. Moreover, by studying pulsars one can learn about the nuclear reactions and interactions between the elementary particles
under these conditions, which cannot be reproduced in terrestrial laboratories. Particle acceleration in the polar region of the magnetic field results in
$\gamma$ radiation, which is expected to have a high degree of polarization~\cite{Kaspi2004}. Depending on the place where the radiation occurs, the pulsar
emission can be explained in the framework of a polar cap model or an outer cap model. In both models, the emission mechanism is similar, but polarization is
expected to be dissimilar~\cite{McConnell2009}; hence, polarimetry could be used to understand the pulsar's emission mechanism.

\subsection{Gamma Ray Bursts}\label{sec_grbs}
Polarization measurements would also help to understand GRBs. GRBs (see \cite{Meszaros2006} for a review) are short and extremely bright bursts of $\gamma$ rays.
Usually, a short-time (from $10^{-3}$~s to about $10^3$~s) peak of radiation is followed by a long-lasting afterglow. The characteristics of the radiation emitted
during the short-time peak and during the afterglow are different. The number of high-energy photons which may be detected during the short-time burst phase is
expected to be small compared with the one for the long-lived emission. While only about 3\% of the energy emitted during the short-time burst is carried by
high-energy photons with $E>100$~MeV, the high energy photons of the afterglow carry about half of the total emitted energy. Therefore, there is a possibility of
observing polarization of high energy photons during the afterglow. The emission mechanism of GRBs, the magnetic composition, and the geometry and morphology of
GRB jets are still uncertain but can be at least partly revealed in this way. For example, in the case of synchrotron emission, the variation of the observed GRB
polarization with viewing angle depends strongly on the degree of order of the magnetic fields.

It is worth noting that several studies have discussed how the degree of polarization, $P$, depends on the GRB emission mechanisms. In one example, using Monte
Carlo methods Toma $\emph{et al.}$~\cite{Toma2009} showed that the Compton drag model is favored when the degree of polarization $P$$\,>\,$$ 80\%$, and $P\sim
20\%$-$70\%$ concerns the synchrotron radiation with the ordered magnetic fields model. Moreover, studies by  Mundell $\emph{et al.}$~\cite{Mundell2013} and
Lyutikov $\emph{et al.}$~\cite{Lyutikov2003} have proven that polarimetry could assist in revealing the geometry of GRB jets.

\subsection{Active Galactic Nuclei}\label{sec_agns}

Most galaxies contain a supermassive black hole, which can power, via accretion and other related astrophysical processes, a compact very bright source with
strong emission in the optical band, i.e. an AGN. Typically AGNs also emit in the radio band, if jet-like outflows are observed in the optical band; highly
relativistic jets powered by the central black hole could also lead to a $\gamma$ ray emitting AGN, or a blazar. For a recent review see~\cite{Dermer2016}. Radio
quiet AGNs have some hard X-rays/soft $\gamma$ rays in their spectra, and their origin is most likely due to accretion.

For blazars, where the jet dominates, polarization measurements in the X and $\gamma$ ray bands can allow us to constrain the morphology and geometry of the
emitting region, as well as place constraints on the various proposed emission mechanisms. Typically the blazar spectra can be described by two broad peaks: one
in the mm to soft X-ray band, and a second peak in the $\gamma$ ray energy range. High levels of polarization for the photons in the low energy maximum band
indicate that synchrotron radiation is a likely source. The high energy maximum is thought to be due to Inverse Compton (IC) of the same synchrotron electrons.
Depending on the source for the photon field responsible for the IC effect, the resulting radiation can be polarized, and measurement of polarization can be used
to disambiguate between the various ``seed'' photon fields. For a review of the expected polarization signatures the hard X-ray to soft $\gamma$ ray energy for
blazars see~\cite{Krawczynski2012}.

\section{Photon Polarimetry in Astrophysics Research}\label{sec_pol}
Several physical processes such as the photoelectric effect, Thomson scattering, Compton scattering, and electron-positron pair production can be used to measure
photon linear polarization. Polarimeters based on the photoelectric effect and Thomson scattering are used at very low energies. Compton polarimeters  are
commonly used  for energies from  50~keV to a few MeV~\cite{Lei1997, McConnell2004, McConnell2006, Krawczynski2011, Hajdas2013, Pohl2014}. Those polarimeters are
not efficient at a photon energy of 100~MeV because kinematical suppression of the Compton rate at large scattering angles leads to a fast drop in the analyzing
power (as 1/E$_\gamma$) above the energy range of a few MeV.

Some of the major achievements in astrophysics that were obtained using polarimetry are: the discovery of synchrotron radiation from the Crab
Nebula~\cite{Oort1956}; the study of the surface composition of solar system objects~\cite{Bowell1974}; the measurement of the X-ray linear polarization of the
Crab Nebula~\cite{Weisskopf1978}, which is still one of the best measurements of linear polarization for astrophysical sources; mapping of solar and stellar
magnetic fields~\cite{Schrijver2000}; detection of polarization in the CMB radiation ~\cite{Kovac2002}; and analysis of large scale galactic magnetic
fields~\cite{Kulsrud2008}.

The measurement of polarization in this high energy $\gamma$ ray regime can be done by detecting the electron-positron pairs produced by $\gamma$ rays and
analysis of a non-uniformity of event distribution in the electron-positron pair plane angle, as discussed in \cite{Kelner1975}. However, implementation of this
technique should consider limitations due to multiple Coulomb scatterings in the detector, and there are no successful polarization measurements for astrophysical
sources in the energy regime of interest in our paper.

A number of missions have included cosmic $\gamma$ ray observations, but only a few of them are capable of measuring polarization. The polarimetry measurements
were mainly restricted to $\gamma$ rays with low energies E $<$ 10~MeV. As an example, the Reuven Ramaty High Energy Solar Spectroscopic Imager (RHESSI)
[hesperia.gsfc.nasa.gov/ rhessi3],launched to image the Sun at energies from 3~keV to 20~MeV, was capable of  polarimetry up to 2~MeV, and the results were
successfully used to study the polarization of numerous solar flares~\cite{McConnell2003, Boggs2006}.

The SPI detector INTErnational Gamma-Ray Astrophysics Laboratory (INTEGRAL) instrument [sci.esa.int/integral] has the capability of detecting polarization in the
range of 3~keV to 8~MeV~\cite{Lei1997, Forot2007}. It was used to measure the polarization of GRB 041219a, and later  a high degree of polarization of $\gamma$
rays from that source~\cite{McGlynn2007, Kalemci2007} was confirmed. The Tracking and Imaging Gamma Ray Experiment (TIGRE) Compton telescope, which observes
$\gamma$ rays in the range of $0.1$ -- $100$~MeV, can measure polarization up to 2~MeV.

Recently, Morselli $\emph{et al.}$ proposed GAMMA-LIGHT to detect $\gamma$ rays in the energy range 10~MeV -- 10~GeV, and they believe that it will provide
solutions to all the current issues that could not be resolved by AGILE and FERMI-LAT in the energy range 10 -- 200~MeV. It can also determine the polarization
for intense sources for energies above a few hundred MeV with high accuracy~\cite{Morselli2013}.

In spite of the limitations due to low sensitivity, there are numerous polarimetry studies in $\gamma$ ray astrophysics, and various proposals have been put forth
regarding medium and high energy $\gamma$ ray polarimeters. In addition to the ones already mentioned in the Introduction, we briefly discuss here some other
important proposals. For example, Bloser $\emph{et al.}$~\cite{Bloser2006} proposed the advanced pair telescope (APT), also a polarimeter, in the $\sim$~50~MeV --
1~GeV range. That proposal uses a gas-based Time Projection Chamber (TPC) with micro-well amplification to track the $e^{+}$, $e^{-}$ path. The polarization
sensitivity was estimated by using Geant4 Monte Carlo simulations. Preliminary results indicated that it will be capable of detecting linearly polarized emissions
from bright sources at 100~MeV. As an updated version of the APT, Hunter $\emph{et al.}$~\cite{Hunter2013} suggested the Advanced Energetic Pair Telescope (AdEPT)
for $\gamma$ ray polarimetry in the medium energy range; further, they mentioned that it would also provide better photon angular resolution than FERMI-LAT in the
range of $\sim$~5 to $\sim$~200~MeV.

HARPO is a hermetic argon TPC detector proposed by Bernard $\emph{et al.}$~\cite{Bernard2012, Bernard2012a} which would have high angular resolution and would be sensitive to
polarization of $\gamma$ rays with energies in the MeV-GeV range. A demonstrator for this TPC was built, and preliminary estimates of the spatial resolution are
encouraging. Currently, the HARPO team is finalizing a demonstrator set up to characterize a beam of polarized $\gamma$ rays in the energy range of 2 --
76~MeV~\cite{Bernard2014}.

\section{The photon flux from the Crab system}\label{sec_flux}

Below we discuss the importance of $\gamma$ ray polarimetry studies for the Crab pulsar/nebula system. After the initial finding of polarization in the $\gamma$
ray flux from the Crab nebula, reported in~\cite{Dean2008}  when analyzing the data from the spectrometer on INTEGRAL, and its subsequent confirmation
in~\cite{Forot2008}, a series of more detailed studies followed, in order to constrain the parameters for the main emission mechanisms. For example, observations
of the $\gamma$ rays from the Crab Pulsar and Crab Nebula have been reported in \cite{Abdo2010} for eight months of survey data with FERMI-LAT. The pulsar
dominates the phase-averaged photon flux, but there is an off-pulse window (the phase interval between $\sim0.5$ and $\sim0.8$ of the pulsar period, lasting about
35\% of the total duration of the cycle) when the pulsar flux is negligible and it is therefore possible to observe the nebular, highly polarized, emission. In
order to determine the polarization fraction and polarization angle for the Crab pulsar/nebula and similar systems, one starts from the azimuthal profile
$N(\psi)$ in Compton counts, which can be parametrized as follows:
\begin{equation}\label{Eq:AzimuthProfile}
N(\psi)=S\left[1+a_0\cos(2\psi-2\psi_0)\right],
\end{equation}
for a source with polarization angle PA = $\psi_0$ and polarization fraction PF =$a_0/a_{100}$. The amplitude $a_{100}$ corresponds to the measured amplitude for
a $100\%$ polarized source. The Crab system is an excellent candidate for studying $\gamma$ ray polarization, and for calibrating any future instruments, since
the phase-averaged PF in the $300-450$keV band is $98\pm37\%$ (see~\cite{Moran2015}), and in the $0.1-1$MeV band the PF is $46\pm10\%$ (see~\cite{Dean2008}).

According to the analysis conducted in~\cite{Abdo2010}, the spectrum of the Crab Nebula in the $100$-$300$~MeV range can be described by the following combined
expression:
\begin{align}
\frac{dN}{dE}=N_{\mathrm{sync}}E^{-\Gamma_{\mathrm{sync}}}+ N_{\mathrm{IC}}E^{-\Gamma_{IC}},
\end{align}
where the quantity $dN/dE$ is measured in $\mathrm{cm}^{-2}\mathrm{s}^{-1}\mathrm{MeV}^{-1}$ representing the number of photons reaching $1$~$\mathrm{cm}^{2}$ of
the detector area per second, per $1$~MeV of energy.
The energy $E$ on the right hand side is measured in GeV.
The prefactors $N_{\mathrm{sync}}\approx9.1\times 10^{-13}\, \mathrm{cm}^{-2}\mathrm{s}^{-1}\mathrm{MeV}^{-1}$ and $N_{\mathrm{IC}}\approx 6.4\times 10^{-12}\,
\mathrm{cm}^{-2}\mathrm{s}^{-1}\mathrm{MeV}^{-1}$ are determined by 35\% of the total duration of the cycle, while $\Gamma_{\mathrm{sync}}\approx4$ and
$\Gamma_{\mathrm{IC}}\approx1.65$.
The first and second terms on the right hand side, as well as the indices "sync" and "IC", correspond to the synchrotron and inverse Compton components of the
spectrum, respectively. As one can see, these terms have different dependence on the energy $E$ since they represent different contributions to the total
spectrum. The first part (the synchrotron radiation) comes from emission by high energy electrons in the nebular magnetic field while the second part is due to
the inverse Compton scattering of the primary accelerated, relativistic, electrons off of the various soft ``seed'' photon fields present such as the synchrotron,
far infrared excess, and cosmic microwave background (see~\cite{Horns2004} for the fourth possible source of ``seed photons''). Polarization measurements could
potentially  be used to disambiguate between those various ``seed'' fields, in a similar manner to the case of blazars discussed in Section~\ref{sec_agns}.

For convenience let us rewrite the expression for the spectrum of the Crab Nebula in the form
\begin{align}
\frac{dN}{dE}=\tilde N_{\mathrm{sync}}E^{-\Gamma_{\mathrm{sync}}}+ \tilde N_{\mathrm{IC}}E^{-\Gamma_{IC}},
\end{align}
where the energy $E$ on both sides is now measured in MeV, therefore $\tilde N_{\mathrm{sync}}\approx9.1\times 10^{-1}\,
\mathrm{cm}^{-2}\mathrm{s}^{-1}\mathrm{MeV}^{-1}$ and $\tilde N_{\mathrm{IC}}\approx 5.7\times 10^{-7}\, \mathrm{cm}^{-2}\mathrm{s}^{-1}\mathrm{MeV}^{-1}$.
Integrating $dN/dE$, for the photon flux above $100$~MeV coming from the Crab Nebula we obtain the number $\sim\,3.5\times10^{-7}\,
\mathrm{cm}^{-2}\mathrm{s}^{-1}$, giving for the total cycle duration $\sim10^{-6}\ \mathrm{cm}^{-2}\mathrm{s}^{-1}$, or $\sim3\times10^5\,
\mathrm{m}^{-2}\mathrm{y}^{-1}$.

At the same time the averaged spectrum of the Crab Pulsar is described in \cite{Abdo2010} as follows:
\begin{align}
\frac{dN}{dE}=N_{0}E^{-\Gamma}\exp\left(-\frac{E}{E_c}\right),
\end{align}
where $N_{0}\approx2.36\times 10^{-4}\ \mathrm{cm}^{-2}\mathrm{s}^{-1}\mathrm{MeV}^{-1}$, $\Gamma\approx2$ and the cut-off energy $E_c\approx 5800\,
\mathrm{MeV}$. As before, the energy $E$ on both sides is measured in MeV. Integrating this expression, for the photon flux above $100$~MeV coming from the Crab
Pulsar we obtain the number $\sim\,2\times10^{-6}\, \mathrm{cm}^{-2}\mathrm{s}^{-1}$, or $\sim6\times10^5\, \mathrm{m}^{-2}\mathrm{y}^{-1}$.
Thus, the Pulsar's photon flux is twice as intensive as the Nebula's. A fast photometer could be added to the polarimeter instrumentation to collect events and
have a temporal tag and consequently distinguish between nebula and pulsar photons, see e.g.~\cite{Meddi2012}.

We will use the numbers above for an estimation of the polarimeter results at a 100~MeV energy cut. For a 500~MeV cut the number of events drops by a factor of five
(because $E_c$ is much higher the exponential factor does not play a role) .

It is worth noting that the estimates following from \cite{Abdo2010} approximately agree with the corresponding estimates made in \cite{Buehler2012} (where the
formulas (1) and (2) describe the synchrotron and inverse Compton components of the Crab Nebula spectrum while the formula (3) describes the averaged Crab Pulsar
spectrum). According to \cite{Buehler2012}, the total Crab Nebula photon flux above $100$~MeV is $\sim\,7.2\times10^{-7}\ \mathrm{cm}^{-2}\mathrm{s}^{-1}$, while
for the Crab Pulsar the value again reads $\sim\,2\times10^{-6}\ \mathrm{cm}^{-2}\mathrm{s}^{-1}$.

\section{The photon polarimetry with pair production}\label{sec_pair_production}

\begin{figure}[h]
\begin{center}
\includegraphics[width=0.65\textwidth,clip=true]{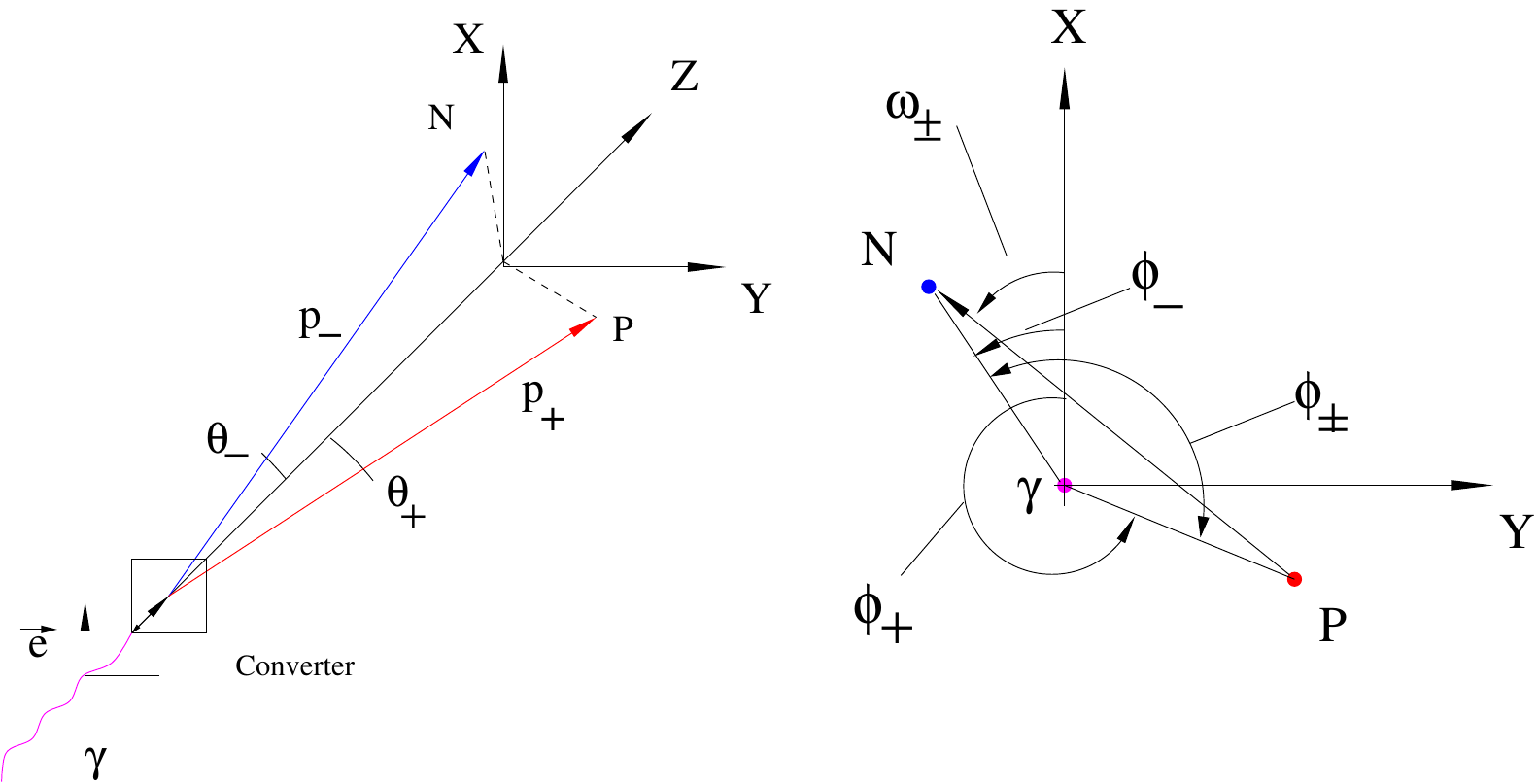}
\end{center}
\caption{\protect The kinematics of $e^+e^-$ pair photo production (left picture) and the azimuthal angles in the detector plane from~\cite{Bogdan2003}. The
photon momentum is directed along the Z axis. The photon polarization vector $\vec{e}$ is parallel to the X axis. The angle $\phi_+,\phi_-$ is the angle between
the photon polarization plane and the plane constructed by the momentum of the photon and the momentum of the positron (the electron). The angle $\phi_\pm$ is
called the co-planarity angle. The labels P and N indicate the positions of the crossings of the detector plane by the positron and the electron. The azimuthal
angle $\omega_\pm$ between the polarization plane and the vector $\overline{PN}$ is a directly measurable parameter.} \label{fig:kinematics}
\end{figure}

The photo production of an electron-positron pair in the field of nuclei is a well understood process which was calculated in QED with all details including the
effect of photon linear polarization, see e.g.~\cite{Maximon1959}. The kinematics and variables of the reactions are shown in Fig.~\ref{fig:kinematics}. The
distribution of events over an azimuthal angle $\phi_{+(-)}$ of a positron (electron) relative to the direction of an incident photon has the following form: \\
$dN/d\phi_\pm \, \propto \, 1 \,+\, A\cdot P_\gamma \cdot \cos 2 (\phi_{+(-)} + \Delta) $, where $A$ is the analyzing power, $P_\gamma$ is the degree of the
photon linear polarization, and $\Delta$ is the angle of the photon linear polarization vector in the detector coordinate system. In practice~\cite{Bogdan2003},
angle $\omega_\pm$ could be used instead of $\phi_{+(-)}$ because at the photon energies of interest the co-planarity angle $\phi_\pm \sim 180^\circ$.


\begin{figure}[htb]
\begin{center}
    \includegraphics[trim = 20mm 0mm 20mm 0mm, width=0.65\textwidth, clip=true]{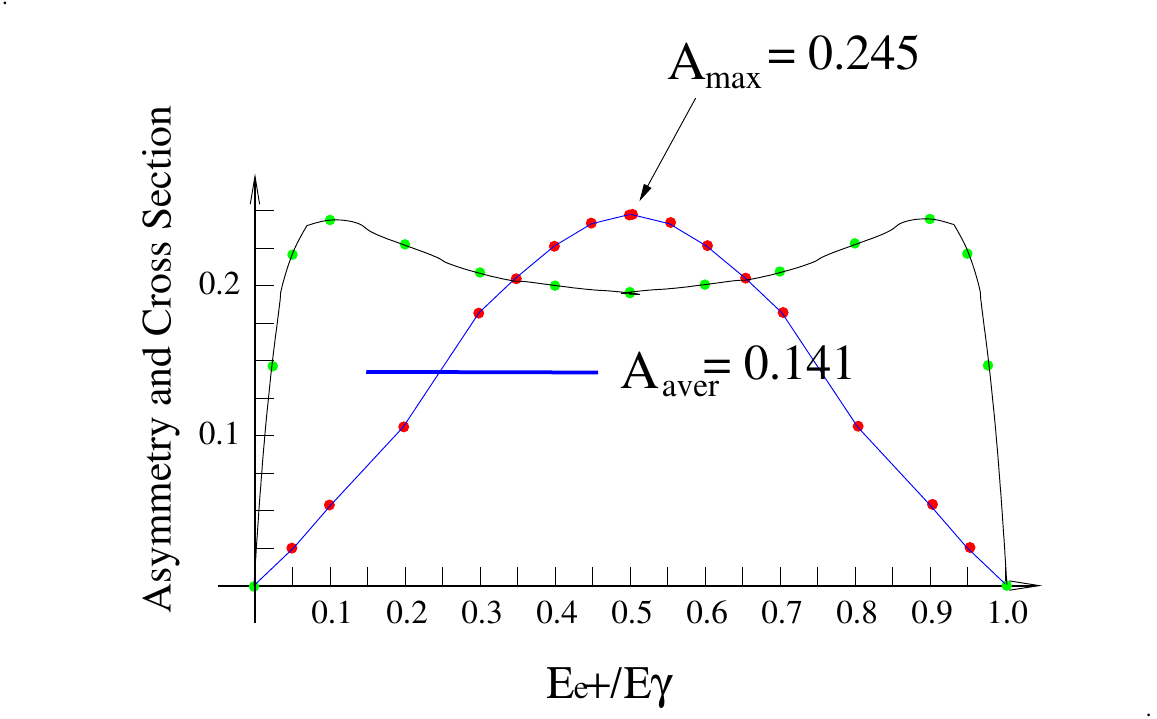}
\end{center}
    \caption{The asymmetry (red dots) and cross section (green dots) of pair production by 2 GeV 100\%
linearly polarized photons as a function of positron energy~\cite{Bogdan2003}. $A_{aver}$ shows the asymmetry averaged over the full range of the positron
energy.}
    \label{fig:asy_e+e-M}
\end{figure}

The value of analyzing power $A$ was found to be a complicated function of the event parameters and detection arrangement~\cite{Maximon1959}. The numerical
integration of the full expression could be performed for given conditions, see e.g.~\cite{Bogdan2003}. In a high energy limit a compact expression for the
integrated analyzing power for the pair photo-production from atomic electron was obtained in~\cite{Boldyshev1971}.

The practical design and the test of the polarimeter for a beam of high energy photons were reported in~\cite{Jager2004}. There we detected both particles of the
pair and reconstructed the azimuthal angle of the pair plane $\omega_\pm$ (see Fig.~\ref{fig:kinematics}). The analyzing power, averaged over energy sharing
between electron and positron and pair open angle of the experimental acceptance, has been found to be 0.116$\pm$0.002, comparable to a 0.14 value as shown in
Fig.~\ref{fig:asy_e+e-M} reproduced from~\cite{Bogdan2003}.
When the pair components move through the converter, the azimuthal angle built on pair coordinates and pair vertex becomes blurred due to multiple scattering.

It is useful to note that the purpose of development in~\cite{Jager2004} was a polarimeter for an intense photon beam. The thickness of the converter in the beam
polarimeter was chosen to be very small to minimize systematics of the measurement of the photon polarization degree. However, the polarimeter could be calibrated
by using the highly polarized photon beams produced in the laser-backscattering facilities. For the cosmic ray polarimeter, we propose a larger converter
thickness and calibration of the device. Such an approach is more productive for cosmic rays studies where a relative systematic error on the polarization degree
at the level of 3-5\% is acceptable.

Let us also note that for the photon beam polarimetry there are additional options such as a coherent pair production in an oriented crystal and a magnetic
separation of the pair components used many years ago in nuclear physics experiments. For the space-borne photon investigation, those polarimeters are not
applicable for the obvious reasons of the limited angle range for the coherent effects and the large weight and power consumption of the magnetic system.

An active converter with a coordinate resolution of a few microns would allow us to construct a dream device, a very efficient polarimeter. A real world
active-converter device, a gas-filled TPC, has a spatial resolution of 100~$\mu$m and much larger two-track resolution of 1.5-2~mm (for a few cm long drift
distance). Such a polarimeter will be a very productive instrument for the photon energy range below 50~MeV. However, because of these resolutions, it would be
hard to measure the degree of polarization of photons whose energy is greater than 100~MeV.

A polarimeter with a separation of the converter and pair detector functions could benefit from the high coordinate resolution of the silicon MSD of 10-15~$\mu$m,
its two-track resolution of 0.2~mm, and flexibility for the distance between a converter and pair hits detector: Between them would be a vacuum gap.

\section{A polarimeter for cosmic {\large $\gamma$} rays}\label{sec_polarimeter}

The key parameters of the polarimeter are the efficiency, $\epsilon$, and analyzing power, $A$. Here we outline the analysis of the Figure-of-Merit, $FOM \,=\,
\epsilon \times A^2$. We will consider a polarimeter as a stack of individual flat cells, each of which is composed of a converter with a two-dimensional
coordinate readout and a coordinate detector for two-track events with no material between them.

The thickness of the converter, where the photon produces the electron-positron pair, defines in the first approximation the polarimeter efficiency as follows:
\begin{equation}
\epsilon \,=\, \eta_{cell} \times \frac{1-r^n}{1-r}
\label{eq:eff}
\end{equation}
The efficiency of one cell is $\eta_{cell} \,=\, 1 - \exp{(-\frac{7}{9}  \times t_{conv})}$, where $t_{conv}$ is the thickness of the converter in units of
radiation length. The $r$ is the reduction of the photon flux due to absorption in a single cell defined as $r \,=\, \exp{(-\frac{7}{9}  \times t_{cell-r})}$,
where $t_{cell-r}$ is the thickness of the cell in units of radiation length, and $n = L/t_{cell-g}$ is the number of cells in the device of length $L$ and
geometrical thickness of the cell $t_{cell-g}$.

The converter thickness needs to be optimized because above some thickness it does not improve the $FOM$ or the accuracy of the polarimeter result (see the next
section). As it is shown in Fig.~\ref{fig:asy_e+e-M}, selection of the symmetric pairs ($E_+ \,=\,  E_-$) provides an analyzing power $A_{sim} \,=\, 0.25$ while
the analyzing power averaged over pair energy sharing $A_{aver} \,=\, 0.14$. However, the value of the $FOM$ is largest when the cut on pair energy sharing is relaxed. The
practical case for the energy cut is $E_+, E_- > E_\gamma/4$, which allows us to avoid events with a low value of $A$ and most $\delta$-electron contamination.
The average value of $A$ for such a range of $e^+-e^-$ energy sharing is 0.20.

The energy of the particles and the shower coordinates could be measured by a segmented electromagnetic calorimeter or estimated from the width of the track,
which due to multiple scattering is inversely proportional to the particle momentum. In the photon energy range of interest, both electron and positron will pass
through a large number of cells. Determination of particle energy based on multiple scattering would provide $\sim$20\% relative energy resolution, which is
sufficient for the proposed cut $E_+, E_- > E_\gamma/4$. Estimation of the particle energy could also be useful for rejection of the hits in MSDs induced by the
$\delta$-electrons.

\section{Monte Carlo simulation}\label{sec_MC}

A Monte Carlo simulation was used to evaluate the general effects of pair production
in the converter and the specific design of the polarimeter.

\subsection{Study of the converter thickness effects}
We used a Geant3-based MC code to study the photon detection efficiency and electron-positron pair azimuthal distribution in a wide range of converter
thickness up to 10\% of radiation length. Because both the pair opening angle and multiple scattering are scaled with the photon energy, the distributions are
almost energy independent. We present first the results for 100~MeV photon energy for different thicknesses of the converter at a fixed distance of 20~mm
between the converter end and the detector.

We used a standard Geant3 pair production generator for the unpolarized photons and at the conversion point introduced a weighting factor for each event as
$W(\omega_\pm) = 1 \,+\, A \times \cos 2 \omega_\pm$ to simulate a polarization effect on an event-by-event basis. The value of azimuthal angle modulation at the
pair production point was fixed at 0.20, which is the average value of the analyzing power $A$ at production over the selected range of particle energies. The
pair component propagation was realized in the MC, and the track parameters were evaluated.

\begin{figure}[thb]
\begin{center}
\includegraphics[trim = 0mm 0mm 0mm 0mm, width=0.65\textwidth,clip=true]{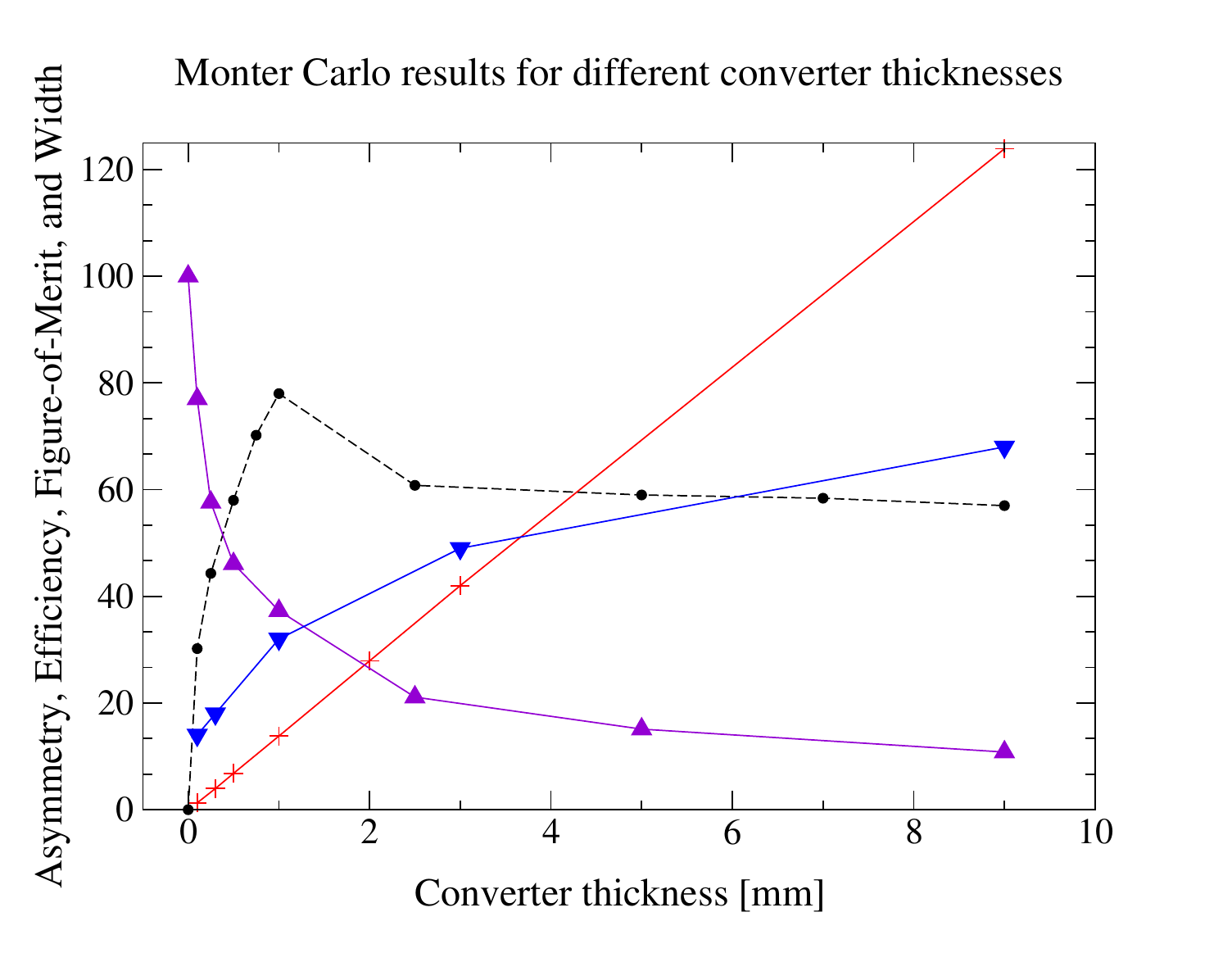}
\end{center}
\caption{\protect Results of MC simulation for 100~MeV photon energy vs. thickness of the converter. The purple set (triangles pointing upward) is for the
analyzing power (normalized to the value 0.20 at the conversion point). The red set (crosses) is the number of $e^+e^-$ pairs with $E_+, E_- > E_\gamma/4$ per $2
\times 10^4$ incident photons. The blue set (triangles pointing downward) is for the rms of the difference $\Delta \omega_\pm$ between the angle $\omega_\pm$ at
the production point and at the detector. The  $\Delta \omega_\pm$ shows a loss of correlation between production and detection points, in units of 10~mrad. The
black set (circles) is the $FOM \,=\, \epsilon \times A^2$ in arbitrary units. A cut $r_{+-} > 0.2$~mm was applied on the distance between coordinates of two
tracks (for 20~mm between the converter and the detector planes).} \label{fig:MC100} \vskip -0.05 in
\end{figure}

Fig.~\ref{fig:MC100} shows the summary of MC results. The apparent optimum converter thickness is close to 1~mm, for which the projected $A = 0.10$ is reasonably
large and the $FOM$ is close to the saturation limit. However, we are expecting that when $\delta$-electron hits are included in analysis the optimum thickness
for the 100~MeV photon case will be smaller and the $FOM$ will be a bit lower.


\subsection{Detector parameters}

The coordinate detector allows determination of the opening angle between the pair components and the azimuthal angle of the pair plane relative to the lab
coordinate system, the main variable for measurement of the photon polarization. Such a detector is characterized by the coordinate resolution, $\sigma_x$, and
the minimum the two-track distance, $a_{x, min}$, at which coordinates of two tracks could be determined with quoted $\sigma_x$ accuracy. The $a_{x, min}$ is
typically 2~mm for the drift chamber. For TPC with a micromega amplification stage and a strip-type readout, $a_{x, min}$ is about 4 strips or 1.6~mm (pitch equal to
400~$\mu m$). For silicon MSD $a_{x, min}$ is about 0.20 mm (pitch equal to 50~$\mu m$).

The opening angle between the pair components is on the order of $4/(E_\gamma/m_e)$, where $E_\gamma$ is the photon energy and $m_e$ is the electron rest mass.
The events with an opening angle larger than $3/(E_\gamma/m_e)$ but less than $9/(E\gamma/m_e)$ provide most of the analyzing power, as it is shown
in~\cite{Bogdan2003}. It is easy to find that the resulting geometrical thickness of the cell is $t_{cell-g} \,=\, (a_{x, min} \cdot E_\gamma)/(3 \cdot m_e)$,
whose numerical values are shown in Tab.~\ref{tab:cell_thickness}.


\begin{table}[htb]\begin{center}
\caption{The geometrical cell thickness, $t_{cell-g}$, in {\bf {cm}} for the different detectors and photon energies.} \vskip 0.1 in
{\begin{tabular}{|c|c|c|c|} 
\hline

Photon energy [MeV] & Drift chamber & TPC/micromega & silicon MSD \\  \hline
20      & 2.6  &    2          &   0.26      \\  \hline

100    & 13   &    10        &   1.3        \\  \hline

500    & 65   &    52        &   6.5     \\  \hline

2000  & 260 &    210      &   26        \\  \hline
\end{tabular}
\label{tab:cell_thickness}}\end{center}
\end{table}

For photon energies above 100~MeV, the silicon MSD is a preferable option because of the limit on the apparatus's total length.
Indeed, considering a 300~cm total length and an energy of 500~MeV, the number of cells is 5 for the drift chamber option,
6 for the TPC/micromega, and 46 for the MSD option.
At the same time, the total amount of matter in the polarimeter should be limited to one radiation length or less, because of significant absorption of
the incident photons which will reduce the average efficiency per cell.
For example, in the MSD option 54\% absorption will occur with 46 cells (1~mm thickness 2d readout converter detector and
two 0.3~mm thickness 1d readout track detectors).
The detection efficiency (pair production in the converters) could be estimated as $\epsilon \,=\, \eta_{cell}
\times \frac{1-r^n}{1-r}$ (Eq.~\ref{eq:eff}), which is about 34\% for the selected parameters of the polarimeter.
However, the useful statistics result is lower due to a cut on the pair components' energies and contributions of the non-pair production processes,
especially at low photon energy.
For a photon energy of 100~MeV, the obtained efficiency is 0.28\% per cell and the overall efficiency for the 30 cell polarimeter is 9\%.
The efficiency becomes significantly larger ($\sim$15\%) for a photon energy of 800 MeV.

Assuming observation of the Crab pulsar photon source with a 1~m$^2$ detector for one year, the total statistics of pairs (above a 100~MeV photon energy cut) was
estimated to be  $N_{pairs} \,=\, 6 \times 10^5 \times 0.09 \,=\, 0.54 \times10^5$.
For the projected analyzing power $A$ of 0.10,  the statistical accuracy of the polarization measurement is
$\sigma_{_P} = \frac{1}{A} \times \sqrt{\frac{2}{N_{pairs}}} \approx 0.10$ for the MSD detector option.
Realization of such high accuracy would require a prior calibration of the polarimeter at a laser-back scattering facility.

\subsection{Geometry of the MSD-based polarimeter}\label{sec_MSD-based}
\begin{figure}[thb]
\begin{center}
\includegraphics[trim = 5mm 5mm 5mm 5mm, width=0.65\textwidth,clip=true]{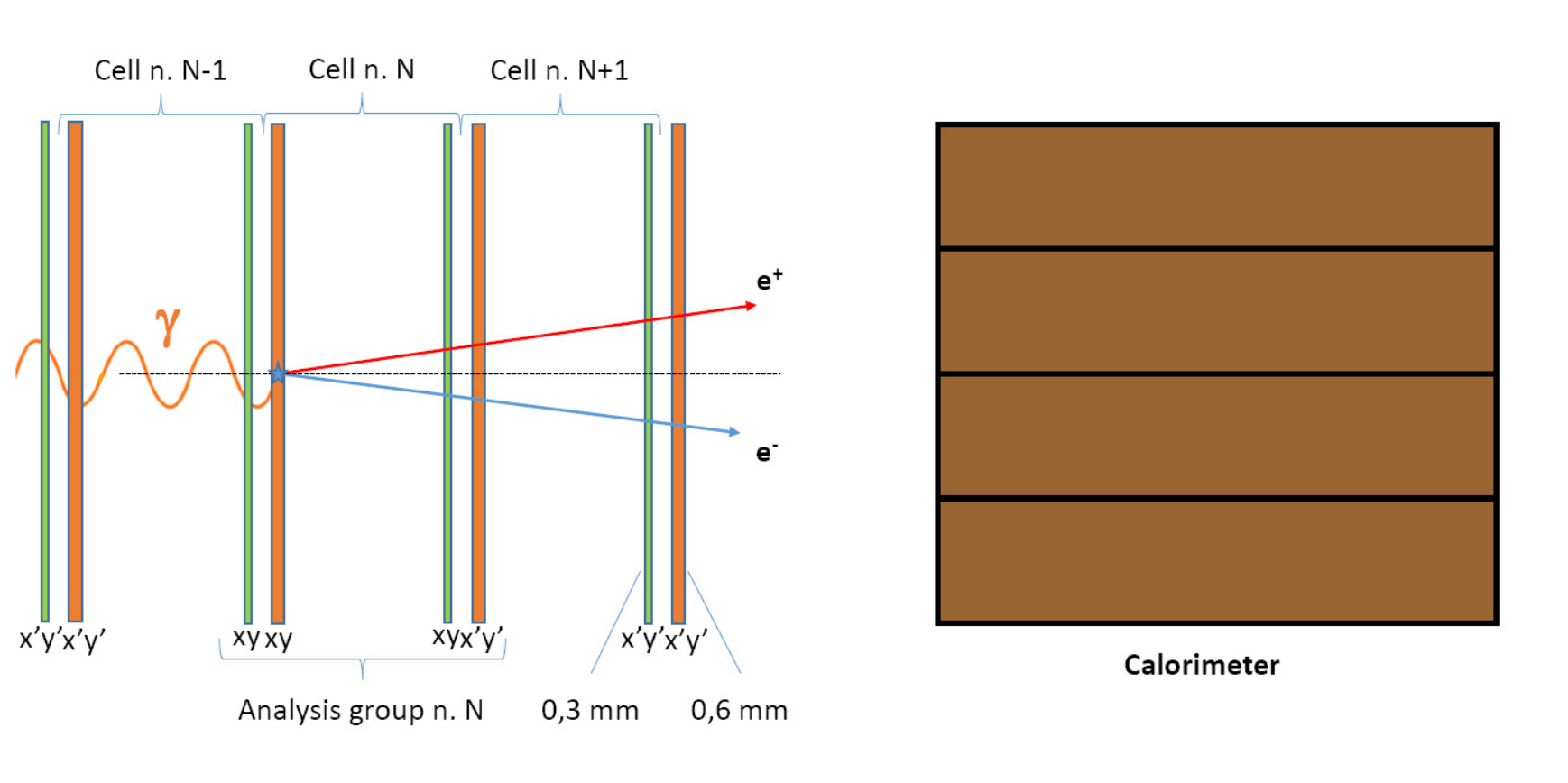}
\end{center}
\caption{\protect A schematic 2D view of three cells and of the calorimeter of the polarimeter.} \label{fig:cells}
\end{figure}

{The polarimeter is composed of thirty cells.
A single cell of the MSD-based polarimeter is composed of two double-sided 1~m x 1~m wide micro-strip detector
planes each separated by 20~mm (see Fig.~\ref{fig:cells}).
The first plane is 0.3~mm thick and the second is 0.6~mm thick, and they both act as a converter.
They determine the coordinates of the gamma annihilation point.
In analysis we are using an analysis group which includes four planes.
The third and fourth planes of the analysis group are components of adjacent cells.
Every other cell has its readout strips rotated by 45$^\circ$ with respect to the previous cell planes.
As a consequence, one cell and the next in line measure two different sets of coordinates ($(X,Y)$ and $(X\prime,Y\prime)$, respectively)}.
Measurement of two sets of coordinates allows unambiguous determination of the two-particle event geometry.


\begin{figure}[thb]
\begin{center}
\includegraphics[trim = 0mm 0mm  0mm  0mm, width=0.65\textwidth,clip=true]{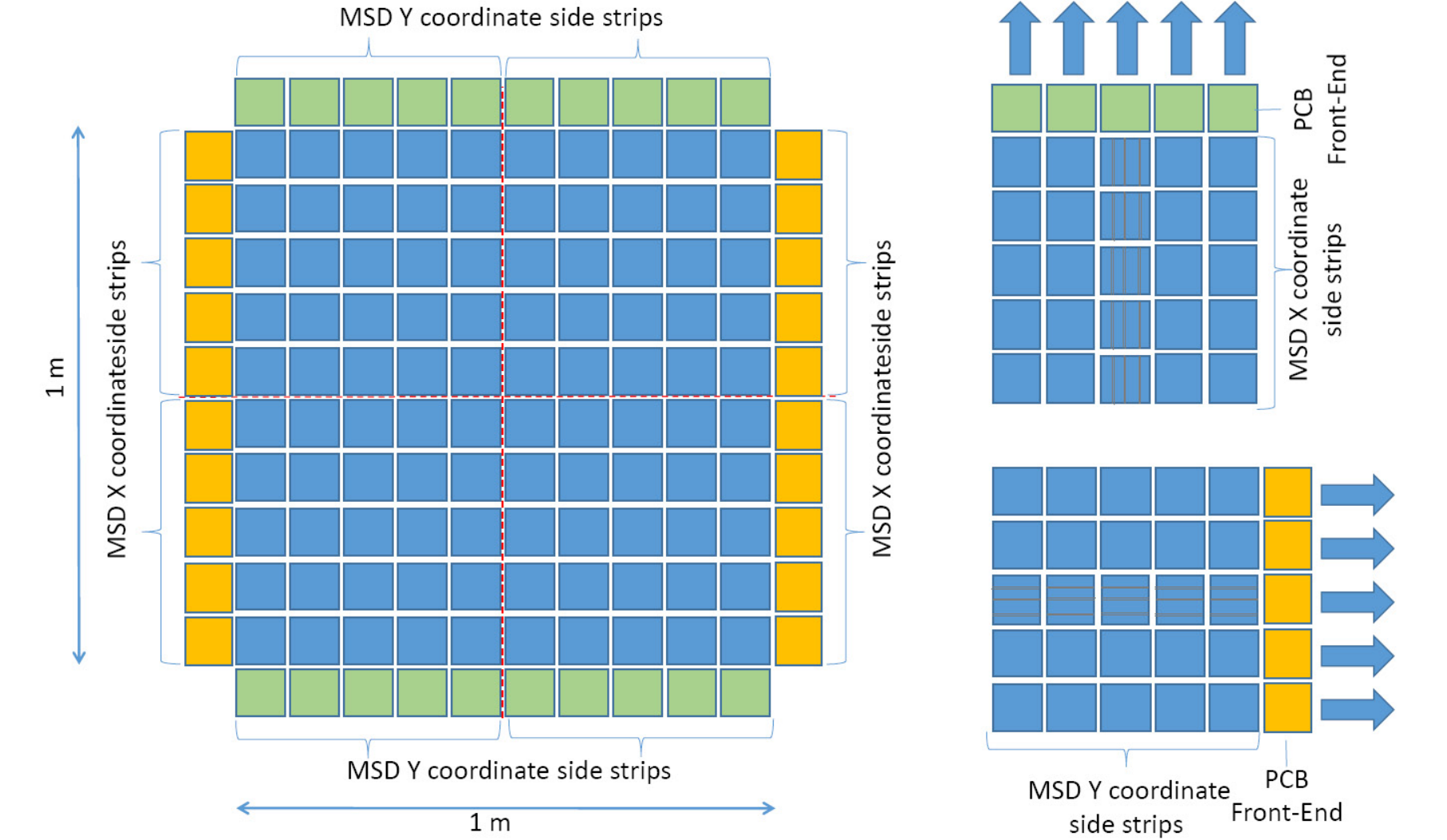}
\end{center}
\caption{Left: the schematic front view of a plane made up of its four quadrants, and their read out electronics. In green, the electronics Printed Circuit Boards
(PCB) of the X coordinate side strips. In yellow, the electronics PCB of the Y coordinate side strips. Upper right: front view of a single quadrant X coordinate
side. Lower right: front view of a single quadrant Y coordinate side. Each quadrant is divided into five rows, each formed by connecting five Micro-Strip
Detectors in a line.} \label{fig:subplane}
\end{figure}

\vskip -0.15 in To avoid strips that are too long and hence a Signal/Noise ratio that is too small, each plane is subdivided into four quadrants, electronically
independent of each other, made up in turn of a mosaic of 5~x~5 DC coupled 100 $\mu$m pitch silicon micro-strip detectors of dimensions 10cm~x~10cm (see
Fig.~\ref{fig:subplane}).
On the side of each plane, the strips of the five silicon micro-strip detectors forming a row (column) of a quadrant are connected in a series
to form five 50~cm long ladders that detect the $X$/$X\prime$ ($Y$/$Y\prime$) track coordinates respectively.
The quadrants are glued to thin kapton foils which electrically insulate them from a carbon-fiber backbone whose thickness is about 0.5~cm (see Fig.~\ref{fig:subplane}).
A one side single micro-strip detector ladders of this kind as long as 72~cm have already been built by~\cite{Barichello1998}.


\begin{figure}[thb]
\begin{center}
\includegraphics[trim = 5mm 5mm 5mm 5mm, width=0.65\textwidth,clip=true]{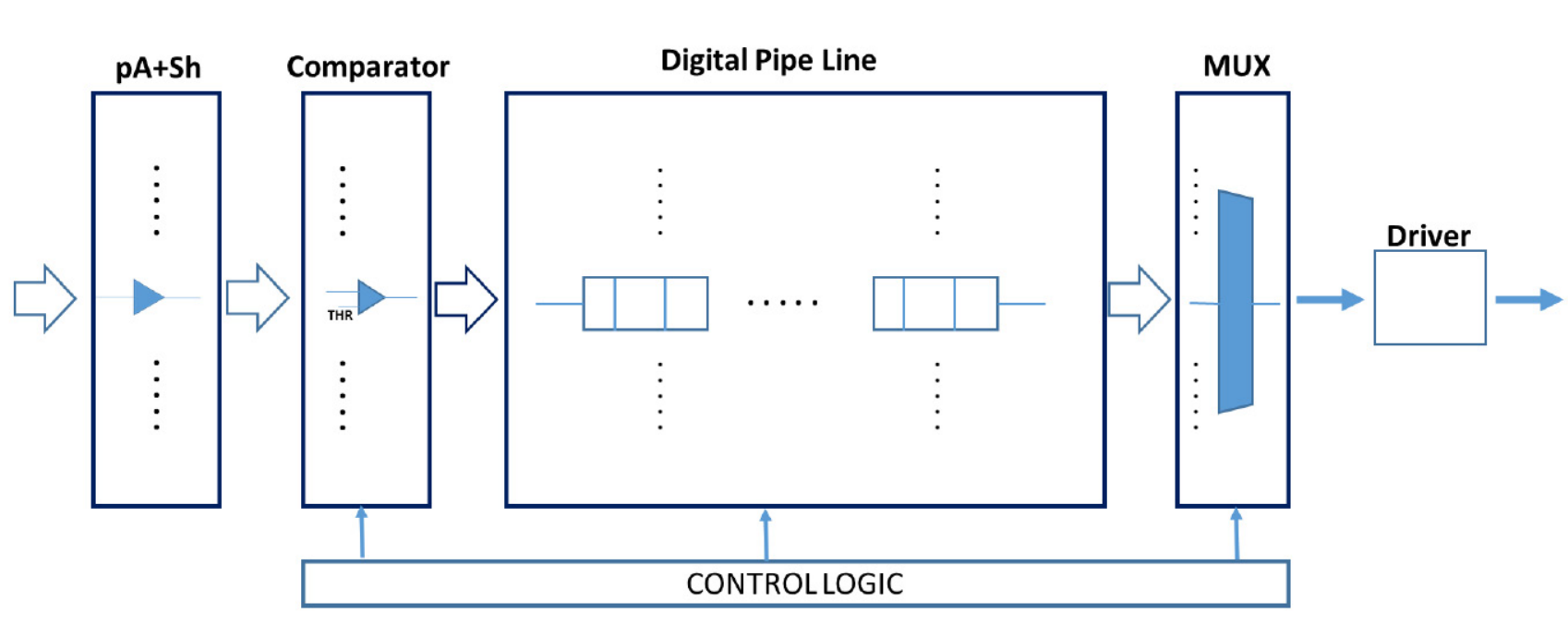}
\end{center}
\caption{The layout of the readout electronics.} \label{fig:readout}  \vskip -.15 in
\end{figure}

To avoid excessive readout electronic power consumption, the analog signals from each 50~cm long strip, after being amplified, discriminated by a comparator, and,
after being digitally pipelined and encoded, are multiplexed in 128 bit words (see Fig.~\ref{fig:readout}). We calculated a power consumption reduction, with
respect to a standard corresponding analog readout, by a factor of six, at least~\cite{Raymond2008}. Considering a standard front end analog readout power
consumption of about 1.2~mW/channel and a figure of 2,480.000 channels for the polarimeter, this adds up to a power consumption of about 450~W. This value will
likely be reduced in the future by the present steady improvement of ASIC (Application-Specific Integrated Circuit) technology.
%
%

A calorimeter will be used for crude ($\sim$20\%) measurement of the photon energy (combined energy of the $e^+e^-$ pair).

\section{Projected results and Conclusions}\label{sec_conclusions}

\begin{figure}[!h]
\begin{center}
\includegraphics[trim = 5mm 5mm 5mm 5mm, width=0.65\textwidth,clip=true]{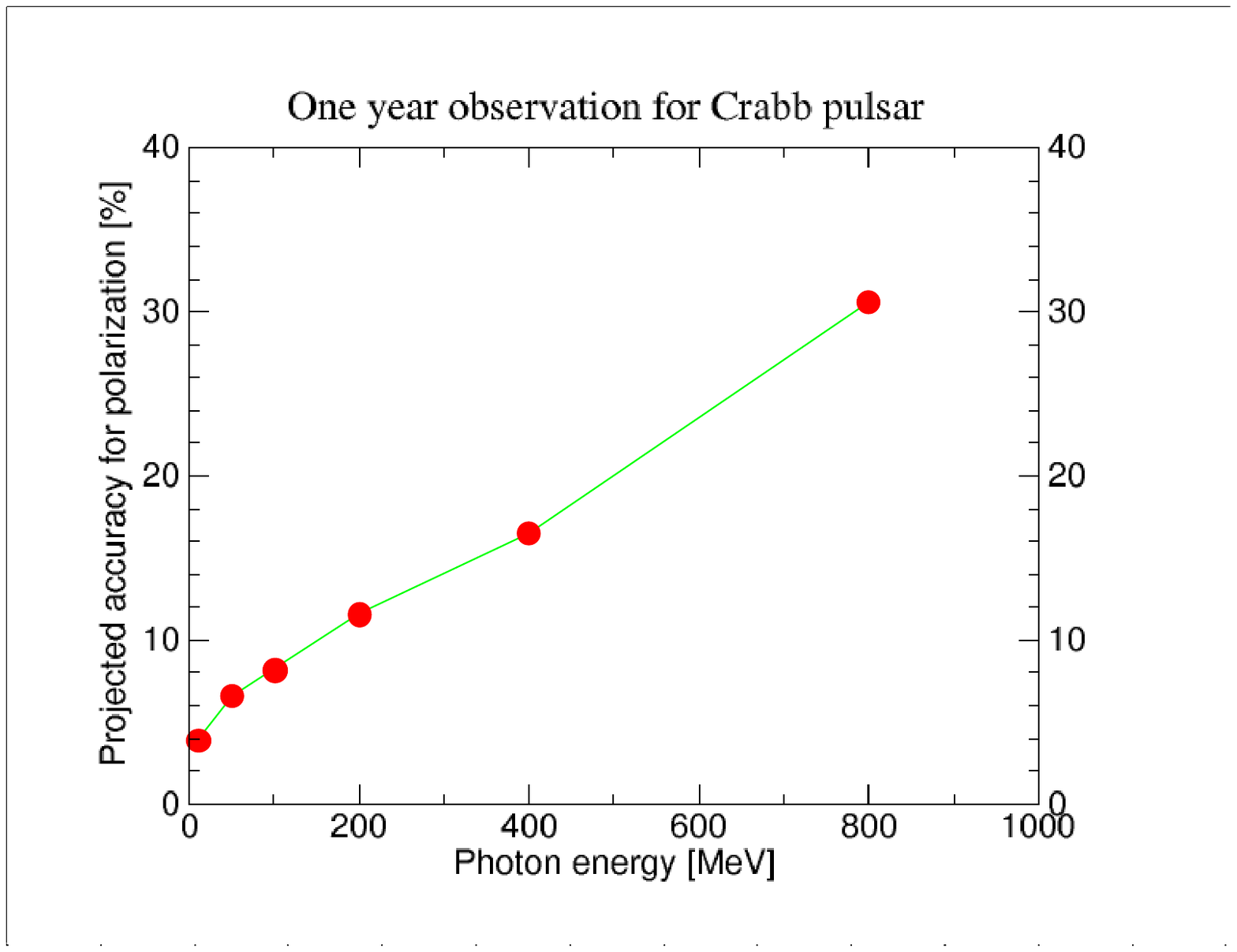}
\end{center}
\caption{Projected accuracy of the polarization measurement with the proposed polarimeter for a different level of photon energy cut.}
\label{fig:projected-result} \vskip -.15 in
\end{figure}

The projected accuracy of polarization measurement is shown in Fig.~\ref{fig:projected-result}.
For each data point we averaged the incident flux over all energies above the shown value.
The photon angular resolution of the proposed system was obtained from MC simulation
as  $\sim$5~mrad for 200~MeV photon energy.


A $\gamma$ ray polarimeter for astrophysics could be constructed using silicon MSD technology.
Each of 30 cells will include one MSD with 2d readout of 0.6 mm thickness and one MSD with 2D readout of 0.3~mm thickness and 0.1~mm pitch.
Using a total of 60~m$^2$ area of MSD (30 cells) the polarimeter would be a device with 9-15\% photon efficiency and an average analyzing power of 10\%..
In a year-long observation, the polarization of the photons from the Crab pulsar would be measured to 8\% accuracy at an energy above
100~MeV and $\sim$30\% accuracy at an energy above 800~MeV.

\section*{Acknowledgements}
\noindent 
This work is supported by NASA award NNX09AV07A and NSF CREST award HRD-1345219.

\end{document}